\documentclass[english]{extarticle}
\usepackage{mathptmx}

\usepackage[T1]{fontenc}
\usepackage[latin9]{inputenc}
\usepackage{geometry}
\usepackage{color}
\usepackage{babel}
\usepackage{float}
\usepackage{rotfloat}
\usepackage{mathrsfs}
\usepackage{amsmath}
\usepackage{amssymb}
\usepackage{setspace}
\usepackage[unicode=true,pdfusetitle,
 bookmarks=true,bookmarksnumbered=false,bookmarksopen=false,
 breaklinks=false,pdfborder={0 0 1},backref=false,colorlinks=false]
 {hyperref}

\makeatletter

\providecommand{\tabularnewline}{\\}

\newcommand{\lyxaddress}[1]{
	\par {\raggedright #1
	\vspace{1.4em}
	\noindent\par}
}

\@ifundefined{date}{}{\date{}}
\makeatother

\begin{document}
\title{\textbf{Generalized quaternionic free rotational Dirac equation and
spinor solutions in the electromagnetic field}}
\author{\textbf{Shikha Bhatt} \textbf{and B. C. Chanyal}\thanks{Corresponding Author (Email: bcchanyal@gmail.com) }}
\maketitle

\lyxaddress{\begin{center}
\textit{Department of Physics, }\\
\textit{G. B. Pant University of Agriculture and Technology, }\\
\textit{Pantnagar-263145, Uttarakhand, India}\\
\par\end{center}}
\begin{abstract}
Starting with the quaternionic Minkowski space-time and its four-vector
representation, a rotational analogue of the quaternionic Dirac equation
in the electromagnetic field is developed, which includes not only
the energy solutions but also the angular momentum solutions for rotating
Dirac fermions. The striking feature of the quaternionic approach
is that it depicts a unified representation of energy-momentum in
a single framework as the space-time. Furthermore, we establish the
generalized Schrodinger-Pauli-like equation associated with generalized
electric and magnetic dipole energies that corresponding to their
dipole moments. As such, the present analysis deals with the invariant behavior of
the extended quaternionic rotational Dirac equation under Lorentz-Poincaré,
Gauge, duality, and CPT invariance.


\end{abstract}

\section{Introduction}

The predictions of the Classical theory have a substantial impact
on the complete description of matter, but its failure to characterize
subatomic particle behavior led to the development of quantum mechanics.
Quantum mechanics employs the idea of a wave function, which has mathematical
construct that provides information in the form of probability amplitudes
\cite{key-1}. With the generalization of de Broglie postulates, Erwin
Schrodinger proposed the non relativistic wave equation, however it
could not be applied to particles traveling with relativistic velocity\textbf{
}\cite{key-2}. Furthermore, by combining quantum mechanics with special
theory of relativity, Dirac and Klein-Gordon discovered the relativistic
wave equation, which became the foundation of `Relativistic Quantum
Mechanics'. The behavior of particles at high energies is described
by relativistic wave equations, which are well invariant under Lorentz
translation. Dirac suggested his renowned differential equation to
correct for the deficiencies of the Klein-Gordon equation, which describes
the motion of spin 1/2 particles \cite{key-3}. The presence of anti-particles
was predicted by the Dirac equation. His equation was expressed in
terms of Dirac matrices, which were also used to describe the interaction
of Dirac particles with an external electromagnetic field. Generally,
researchers used a four-vector potential with scalar and vector potentials
to characterize an external electromagnetic field. In this case, the
Dirac equation can be changed using minimal coupling when a particle
interacts with an electromagnetic field \cite{key-4}. The hyper-complex
algebra is now more commonly used in modern physics to express many
physical quantities. The first hyper-complex algebra that is commonly
used in physics is the quaternion algebra \cite{key-5}. According
to Dickson \cite{key-6}, there are four forms of division algebra:
real numbers, complex numbers, quaternion, and octonion algebra. Complex
numbers (two dimensions), quaternionic algebra (four dimensions),
and octonion algebra (eight dimensions) are all extensions of real
numbers. Many novel results in modern physics have been explained
using hypercomplex algebras \cite{key-7,key-8,key-9,key-10,key-11,key-12,key-13,key-14,key-15,key-16,key-17,key-18,key-19,key-20}.
Quaternions have a structure that is extremely similar to that of
four vectors, with quaternionic counterparts of four vectors such
as four momentum, four velocity, four potential, and so on \cite{key-21}.
Apart from that, the quaternionic algebra has been used to describe
a variety of physical theories \cite{key-22,key-23,key-24,key-25},
including Maxwell's electromagnetic field equations, Dirac Lagrangian,
dual magneto-hydrodynamics (DMHD) equations, and the Bernoulli and
Navier--Stokes equations for dyonic plasma, and so on. Davies \cite{key-26}
examined the Dirac equation in quaternionic form, and Rawat et al.
\cite{key-27} discussed the quaternionic Dirac equation in the bi-spin
state, providing positive and negative energy solutions. Rawat and
Negi \cite{key-28} have proposed a link between the quaternionic
form of the Dirac equation and supersymmetry in presence of electromagnetic
field \cite{key-29}. The non-conservation of the photon energy-momentum
tensor has been investigated \cite{key-30,key-31,key-32,key-33,key-34},
which is more in accordance with the terminology used in this paper.
Non-conservation of the photon energy-momentum tensor \cite{key-34}
is also identified in (i) de Broglie- Proca's theory (ii) Standard-Model
Extension (iii) Non-Linear Electromagnetism. Chanyal \cite{key-35,key-36}
provided a fresh notion for relativistic quantized electromagnetic
fields theory of dyons in terms of quaternion variable. In addition,
the quaternionic algebra has been utilized to represent the rotational
motion of a rigid body \cite{key-37}, where the quaternionic unit
elements are typically associated with matrices of the rotational
group SO(3). Beside, on the $R\times S^{3}$ topology, Carmeli \cite{key-38,key-39,key-40}
studied the Klein-Gordon equation, the Weyl equation, and the Dirac
equation in rotational form. The rotational analogue of the dynamics
of relativistic spin-1/2 fermions and the rotational Dirac equation
in terms of quaternionic form have recently been constructed \cite{key-41}.
So, in light of the aforementioned literature on rotating Dirac fermions,
we have extended the quaternionic rotational Dirac equation in the
presence of an external electromagnetic (EM) field.

Starting with the quaternionic Minkowski space-time and its four-vector
representation, we write the rotational analogue of the generalized
quaternionic Dirac equation for a free fermion spinor where the rotational
Dirac matrices are associated with quaternionic basis vectors. We
also define the fermionic spinor field for quaternions. The energy-momentum
like solutions for Dirac rotating particles are expressed in terms
of four-component spinors. Using the quaternionic minimum coupling
of four-momentum with the EM-field, a rotational analogy of the quaternionic
Dirac equation in the EM-field is developed, which contains not only
the rotational energy (corresponding to the quaternionic scalar component)
but also the angular momentum (corresponding to the quaternionic vector
component) for Dirac fermions associated with electric and magnetic
fields. Furthermore, the solutions for these quaternionic rotational
energy and momentum equations in the EM-field are obtained in terms
of the four component form of the Dirac spinors with spin up and down
state. We establish the generalized Schrodinger-Pauli like energy
equation, which is associated with unperturbed and perturbed energies
due to EM-field interaction. The generalized electric and magnetic
dipole moments are constructed corresponding to generalized electric
and magnetic dipole energies. Moreover, we also addressed the invariant
behavior of the extended quaternionic rotational Dirac equation for
Lorentz, gauge, duality, and CPT invariance to check the various symmetries.

\section{The quaternions}

The quaternions, denoted by $\mathbb{H}$, are a type of hypercomplex
algebra which is a four-dimensional norm division algebra over a set
of real numbers $(\mathbb{R})$. It is a four-dimensional extension
of complex numbers with a scalar and a vector part that is used to
explain dynamic behaviour. A quaternion variable $Q_{q}\in\mathbb{H}$
can be represented as 
\begin{align}
Q_{q}= & \,\,e_{0}x_{0}+\left(e_{1}x_{1}+e_{2}x_{2}+e_{3}x_{3}\right)\,,\label{eq:1}
\end{align}
where ($x_{0},x_{1},x_{2},x_{3})\in\mathbb{R}$ and ($e_{0},e_{1},e_{2},e_{3}$)
are the fundamental quaternionic units known as the quaternionic basis
elements. However, the scalar unit is denoted by $e_{0}$, and the
vector units are denoted by $e_{j}$ $(j=1,2,3)$. The quaternion
multiplication can be determined by the following rules:
\begin{align}
e_{0}^{2}= & \thinspace\,\,e_{0}=-e_{k}^{2}=1\,\,,\nonumber \\
e_{0}e_{k}= & \,\,\,e_{k}e_{0}=e_{k}\thinspace\thinspace,\nonumber \\
e_{k}e_{l}= & \,-\delta_{kl}e_{0}+\epsilon_{klm}e_{m}\,\thinspace,\thinspace\,\,\,\,\forall(k,l,m=1,2,3)\,\thinspace.\label{eq:2}
\end{align}

\noindent Here $\delta_{kl}$ represents delta symbol and $\epsilon_{klm}$
represents the Levi-Civita symbol. Moreover, the quaternionic multiplication
(\ref{eq:2}) can also be written as given Table-1. 
\begin{table}[t]
\begin{centering}
\begin{tabular}{c|cccc}
\hline 
 & $e_{0}$ & $e_{1}$ & $e_{2}$ & $e_{3}$\tabularnewline
\hline 
$e_{0}$ & $1$ & $e_{1}$ & $e_{2}$ & $e_{3}$\tabularnewline
$e_{1}$ & $e_{1}$ & $-1$ & $e_{3}$ & $-e_{2}$\tabularnewline
$e_{2}$ & $e_{2}$ & $-e_{3}$ & $-1$ & $e_{1}$\tabularnewline
$e_{3}$ & $e_{3}$ & $e_{2}$ & $-e_{1}$ & $-1$\tabularnewline
\hline 
\end{tabular}
\par\end{centering}
\caption{Quaternion multiplication table}

\end{table}
 As such, the relation
\begin{align}
e_{k}e_{l}+e_{l}e_{k}=\,\thinspace & -2\delta_{kl}\,\thinspace,\nonumber \\
e_{k}e_{l}-e_{l}e_{k}=\,\thinspace & 2\epsilon_{klm}e_{m}\,\,,\label{eq:3}
\end{align}

\noindent represent the anti-commutation and commutation property
of the quaternionic units. The addition of two quaternions will be
expressed as
\begin{align}
Q_{q}+R_{q}= & \,\thinspace e_{0}\left(x_{0}+y_{0}\right)+e_{1}\left(x_{1}+y_{1}\right)+e_{2}\left(x_{2}+y_{2}\right)+e_{3}\left(x_{3}+y_{3}\right)\nonumber \\
= & \,\thinspace e_{0}z_{0}+\left(e_{1}z_{1}+e_{2}z_{2}+e_{3}z_{3}\right)\equiv T_{q}\thinspace\,,\thinspace\thinspace\thinspace\:\,\left(Q_{q},R_{q},T_{q}\thinspace\in\thinspace\mathbb{H}\right)\thinspace\thinspace,\label{eq:4}
\end{align}
which satisfies the closure property of addition. As such, the quaternions
show the commutative and associative properties of addition, respectively,
$Q_{q}+R_{q}=R_{q}+Q_{q}$ and $\left(Q_{q}+R_{q}\right)+T_{q}=Q_{q}+\left(R_{q}+T_{q}\right).$
The quaternionic multiplication of any two quaternions can be written
by using Table-1 as 
\begin{align}
Q_{q}\circ R_{q}= & \,\thinspace(e_{0}x_{0}+e_{1}x_{1}+e_{2}x_{2}+e_{3}x_{3})\circ(e_{0}y_{0}+e_{1}y_{1}+e_{2}y_{2}+e_{3}y_{3})\nonumber \\
= & \,\,e_{0}\left(x_{0}y_{0}-\overrightarrow{x}\cdot\overrightarrow{y}\right)+e_{j}\left(x_{0}y_{j}+x_{j}y_{0}+\left(\overrightarrow{x}\times\overrightarrow{y}\right)_{j}\right)\,\,,\,\,\,\left(\forall\,\ensuremath{j=1,2,3}\right),\nonumber \\
= & \,\,\text{Sr}\{Q_{q}\circ R_{q}\}+\text{Vr}\{Q_{q}\circ R_{q}\}\label{eq:5}
\end{align}

\noindent where `$\circ$' indicates the quaternion product while
$(\cdot)$ and $(\times)$ are the usual scalar and vector products.
The scalar part of the quaternionic multiplication is `$\text{Sr}$',
while the vector part is `$\text{Vr}$'. Under multiplication, quaternions
are associative but not commutative since $\overset{\rightarrow}{x}\times\overset{\rightarrow}{y}\neq\overset{\rightarrow}{y}\times\overset{\rightarrow}{x}.$
The quaternionic conjugate ($Q_{q}^{\ast}$) of $Q_{q}$ for which
$e_{0}^{*}=e_{0},\,e_{1}^{*}=-e_{1},\,e_{2}^{*}=-e_{2},\,e_{3}^{*}=-e_{3}$
is defined as 
\begin{align}
Q_{q}^{\ast}= & \,\thinspace e_{0}x_{0}-\left(e_{1}x_{1}+e_{2}x_{2}+e_{3}x_{3}\right)\,\,.\label{eq:6}
\end{align}
Thus, using the conjugation property of quaternion, the real part
(i.e. scalar) can be distinguished from its imaginary part (i.e. vector),
as
\begin{align}
\text{Sr}=\thinspace\frac{1}{2}(\thinspace Q_{q}+Q_{q}^{*})=\thinspace & x_{0}\,\,,\nonumber \\
\text{Vr}=\thinspace\frac{1}{2}(\thinspace Q_{q}-Q_{q}^{*})=\thinspace & e_{1}x_{1}+e_{2}x_{2}+e_{3}x_{3}\thinspace\thinspace.\label{eq:7}
\end{align}
Notice that, the quaternions are called real quaternions if the vector
part is zero, and pure quaternions if the scalar part is zero. As
such, the modulus (or norm) of a quaternion is expressed as 

\begin{equation}
\mid Q_{q}\mid=\,\,\sqrt{Q_{q}^{*}\circ Q_{q}}=\,\,\sqrt{x_{0}^{2}+x_{1}^{2}+x_{2}^{2}+x_{3}^{2}}\,\,,\label{eq:8}
\end{equation}

\noindent which satisfy
\begin{equation}
\mid Q_{q}R_{q}\mid\,\thinspace=\thinspace\,\mid Q_{q}\mid\mid R_{q}\mid\,\,.\label{eq:9}
\end{equation}

\noindent The quaternionic inverse becomes
\begin{align}
Q_{q}^{-1}= & \,\,\frac{Q_{q}^{*}}{\mid Q_{q}\mid^{2}}\:\thinspace,\:\thinspace\:\left(\mid Q_{q}\mid\,\neq\,0\right)\,\thinspace.\label{eq:10}
\end{align}
A scalar product can also be defined for quaternions as\textbf{
\begin{align}
\left(Q_{q}\cdot R_{q}\right)= & \,\thinspace-\frac{1}{2}(Q_{q}\circ R_{q}^{*}+R_{q}\circ Q_{q}^{*})=\,\thinspace-\frac{1}{2}(Q_{q}^{*}\circ R_{q}+R_{q}^{*}\circ Q_{q})\,\thinspace.\label{eq:11}
\end{align}
}

\noindent Even so, the quaternionic basis may also be shown as a $2\times2$
matrix form, such that $e_{0}=\,1_{2\times2}$ and $e_{j}=-i\sigma_{j}$
where $\sigma_{j}$ are the standard Pauli matrices \cite{key-29}.
Furthermore, quaternions show a remarkable resemblance to rotational
tau matrices (or isospin matrices) and can be utilized to discuss
particle rotational motion \cite{key-37}.

\section{Quaternionic Minkowski space-time}

\noindent The quaternions can be utilized as an analogous version
of the four vectors because they are regarded comparable to the 4-dimensional
Minkowski space-time with structure $(-,+,+,+)$. Thus, 
\begin{align}
X_{q}=\thinspace\,\left(-ict,\overset{\rightarrow}{R}\right)= & \,\,\left\{ e_{0}\left(-ict\right)+e_{1}X_{1}+e_{2}X_{2}+e_{3}X_{3}\right\} \thinspace,\label{eq:12}\\
X_{q}^{*}=\thinspace\thinspace\left(-ict,-\overset{\rightarrow}{R}\right)= & \,\,\left\{ e_{0}\left(-ict\right)-e_{1}X_{1}-e_{2}X_{2}-e_{3}X_{3}\right\} \thinspace,\label{eq:13}
\end{align}
where $X_{q}$ is the quaternionic 4-position. As such, the quaternionic
4-displacement vector, 4-velocity, and 4-gradient will be expressed
as, respectively,
\begin{align}
ds_{q}=\thinspace\thinspace\left(-icdt,dx,dy,dz\right) & =\,\thinspace e_{0}\left(-icdt\right)+e_{1}dx+e_{2}dy+e_{3}dz\thinspace,\label{eq:14}\\
U_{q}=\,\thinspace\left(-ic,\overset{\rightarrow}{u}\right) & =\thinspace\thinspace e_{0}\left(-ic\right)+e_{1}u_{1}+e_{2}u_{2}+e_{3}u_{3}\,,\label{eq:15}\\
D_{q}=\thinspace\thinspace\left(-\frac{1}{ic}\frac{\partial}{\partial t},\overrightarrow{\nabla}\right) & =\thinspace\thinspace e_{0}\left(-\frac{1}{ic}\frac{\partial}{\partial t}\right)+e_{1}\frac{\partial}{\partial x}+e_{2}\frac{\partial}{\partial y}+e_{3}\frac{\partial}{\partial z}\,.\label{eq:16}
\end{align}

\noindent The quaternionic 4-momentum can be written as 
\begin{align}
P_{q}=\,\thinspace\left(-i\frac{E}{c},\overset{\rightarrow}{p}\right)= & \,\thinspace e_{0}\left(-i\frac{E}{c}\right)+e_{1}p_{1}+e_{2}p_{2}+e_{3}p_{3}\thinspace.\label{eq:17}
\end{align}

\noindent Further, the generalized quaternionic four potential ($V_{q}$),
the electric four potential ($A_{q}$), and the magnetic four potential
($B_{q}$) can all be expressed in terms of Minkowski structure as
\begin{align}
V_{q}=\,\thinspace\left(-i\frac{\Phi}{c},\overset{\rightarrow}{V}\right)= & \,\thinspace e_{0}\left(-i\frac{\Phi}{c}\right)+e_{1}V_{1}+e_{2}V_{2}+e_{3}V_{3}\,,\label{eq:18}\\
A_{q}=\,\,\left(-i\frac{\phi_{e}}{c},\overset{\rightarrow}{A}\right)= & \,\thinspace e_{0}\left(-i\frac{\phi_{e}}{c}\right)+e_{1}A_{1}+e_{2}A_{2}+e_{3}A_{3}\,,\label{eq:19}\\
B_{q}=\thinspace\,\left(-i\frac{\phi_{m}}{c},\overset{\rightarrow}{B}\right)= & \,\thinspace e_{0}\left(-i\frac{\phi_{m}}{c}\right)+e_{1}B_{1}+e_{2}B_{2}+e_{3}B_{3}\,,\label{eq:20}
\end{align}

\noindent where the generalized vector potential ($\overset{\rightarrow}{V}$)
and scalar potential ($\Phi$) are written as $\overset{\rightarrow}{V}=\,\thinspace\overset{\rightarrow}{A}-ic\overset{\rightarrow}{B}$
and $\Phi=\thinspace\,\phi_{e}-ic\phi_{m}$, respectively. Additionally,
the quaternionic mass can be expressed as,\cite{key-42}
\begin{align}
M_{q}= & \,\,e_{0}m_{0}+\left(e_{1}m_{1}+e_{2}m_{2}+e_{3}m_{3}\right)\,\thinspace,\label{eq:21}
\end{align}

\noindent where $m_{0}=\frac{E_{0}}{c^{2}}$ is defined the rest mass
while $m_{j}=\thinspace\left|\frac{p_{j}}{u_{j}}\right|$, $\forall(j=1,2,3)$
is defined the moving mass. The quaternionic moment of inertia (MOI)
can thus be written as
\begin{align}
I_{q}= & \,\thinspace M_{q}\left(X_{q}\circ\left(X_{q}\right)^{*}\right)=M_{q}\left(X_{0}^{2}+X_{1}^{2}+X_{2}^{2}+X_{3}^{2}\right)\nonumber \\
= & \,\thinspace e_{0}I_{0}+\left(e_{1}I_{1}+e_{2}I_{2}+e_{3}I_{3}\right)\thinspace,\label{eq:22}
\end{align}

\noindent where $I_{0}\sim m_{0}|X|^{2}$ is the MOI about the quaternionic
$e_{0}$ axis while $I_{j}\sim m_{j}|X_{j}|^{2}$ are the MOI about
the quaternionic $e_{j}$ axes. We can also express the quaternionic
rotational 4-momentum as using the above quaternionic quantities,
i.e.,
\begin{align}
L_{q}= & \,\,X_{q}\circ P_{q}\,\thinspace=\,\thinspace e_{0}E_{0}+\left(e_{1}L_{1}+e_{2}L_{2}+e_{3}L_{3}\right)\thinspace.\label{eq:23}
\end{align}

\noindent Here, equation (\ref{eq:23}) gives the scalar component
$(E_{0})$ that can be represented as quaternionic rest mass energy,
while the vector components $(\overrightarrow{L})$ can represented
as pure quaternionic angular momentum. So, we have
\begin{align}
E_{0}= & \,\thinspace X_{0}p_{0}-\overset{\rightarrow}{X}\cdot\overset{\rightarrow}{p}\thinspace,\label{eq:24}\\
\overrightarrow{L}= & \,\,X_{0}\overrightarrow{p}+p_{0}\overrightarrow{X}+\left(\overrightarrow{X}\times\overrightarrow{p}\right)\thinspace\,.\label{eq:25}
\end{align}
Interestingly, the quaternionic rest mass energy consisted the scalar
terms while the quaternionic angular momentum consisted all the vector
terms. We get the pure rotational energy $E_{0}\rightarrow-\overrightarrow{X}\cdot\overrightarrow{p}$
and angular momentum $\overrightarrow{L}\rightarrow\left(\overrightarrow{X}\times\overrightarrow{p}\right)$
for a pure quaternion, while for real quaternionic consideration,
we only have the rest mass energy $E_{0}\rightarrow X_{0}p_{0}$ with
no rotational motion $\overrightarrow{L}\rightarrow0$. 

\section{Quaternionic Rotational Dirac (QRD) equation}

A relativistic equation that explains the dynamic behaviour of spin
1/2 particles is the Dirac equation. Let us start with the rotational
analogue of the Dirac equation in order to write the generalized quaternionic
Dirac equation for a free fermion spinor field \cite{key-41}, as
\begin{align}
\left(H_{q}\circ L_{q}-\mathscr{\mathfrak{B}}\lambda^{2}I_{q}\right)\circ\mathtt{\boldsymbol{\Psi}}= & \,\thinspace0\,\thinspace,\label{eq:26}
\end{align}
where $H_{q},L_{q},\mathscr{\mathfrak{B}},\lambda,I_{q}$ and $\boldsymbol{\Psi}$
are rotational analogous to the standard Dirac variables $\overrightarrow{\alpha}$,
$\overrightarrow{p}$, $\beta,\,c,\,m$ and $\psi$. The rotational
Dirac matrix elements $H_{q}$ and $\mathscr{\mathfrak{B}}$ can be
represented in terms of $2\times2$ matrix as 

\noindent 
\begin{align}
H_{q}= & \,\,\,\left\{ D^{0}\left(H\right),D^{j}\left(H\right)\right\} \,,\,\,\,\forall\,D^{0}(H)=\thinspace\left(\begin{array}{cc}
e_{0} & 0\\
0 & e_{0}
\end{array}\right)\,,\,D^{j}(H)=\thinspace\left(\begin{array}{cc}
0 & ie_{j}\\
ie_{j} & 0
\end{array}\right)\,,\label{eq:27}\\
\mathscr{\mathfrak{B}}= & \,\,\left(\begin{array}{cc}
e_{0} & 0\\
0 & -e_{0}
\end{array}\right)\thinspace,\label{eq:28}
\end{align}
where $D^{0}\left(H\right)$ and $D^{j}\left(H\right)$ for $\left(\ensuremath{j=1,2,3}\right)$
are the quaternionic $D$-matrices. Here the speed of rotating fermions
is denoted by `$\lambda$' i.e., $\lambda=\,c\sqrt{\frac{M_{q}}{I_{q}}}=\frac{c}{\mid X\mid}$
(see Ref. \cite{key-40}). Therefore, the generalized QRD equation
becomes
\begin{align}
\left[e_{0}\right. & \left(D^{0}\left(H\right)E_{0}-\lambda\left(\overrightarrow{D}\left(H\right)\cdot\overrightarrow{L}\right)-\mathfrak{B}\lambda^{2}I_{0}\right)\nonumber \\
+ & \left.e_{j}\left(\lambda D^{0}\left(H\right)L_{j}+D^{j}\left(H\right)E_{0}+\lambda\left(\overrightarrow{D}\left(H\right)\times\overrightarrow{L}\right)_{j}-\mathscr{\mathfrak{B}}\lambda^{2}I_{j}\right)\right]\circ\boldsymbol{\Psi}=\,\,0\,\,,\label{eq:29}
\end{align}
The advantage of QRD equation is that it describing not only the rotational
energy but also the rotating momentum for the quaternionic scalar
and vector parts. This is the most significant aspect of quaternionic
analysis, as it allows for dual representation of four vectors in
a single framework. Thus, the energy-momentum solutions of the generalized
QRD equation (\ref{eq:29}) can be obtained with the quaternion fermionic
spinor field $(\boldsymbol{\Psi})$ as\textbf{
\begin{align}
\boldsymbol{\Psi}= & \,\thinspace e_{0}\Psi_{0}+\left(e_{1}\Psi_{1}+e_{2}\Psi_{2}+e_{3}\Psi_{3}\right)\,\thinspace,\nonumber \\
\simeq & \:\left(\begin{array}{c}
\Psi_{0}\\
\Psi_{1}\\
\Psi_{2}\\
\Psi_{3}
\end{array}\right)\,.\label{eq:30}
\end{align}
}

\noindent Equation (\ref{eq:30}) indicates the four components representation
of quaternionic Dirac spinor, and it can be simplified as $\boldsymbol{\Psi}=\left(\Psi_{K}+e_{2}\Psi_{L}\right)$,
where $\Psi_{K}=\,\,\left(\Psi_{0}+e_{1}\Psi_{1}\right)$ and $\Psi_{L}=\,\thinspace\left(\Psi_{2}-e_{1}\Psi_{3}\right)$.
So, the energy solutions of equation (\ref{eq:29}) can be examined
by equating the scalar coefficient ($e_{0}$) :
\begin{align}
\left[D^{0}\left(H\right)E_{0}-\lambda\left(\overrightarrow{D}\left(H\right)\cdot\overrightarrow{L}\right)-\mathfrak{B}\lambda^{2}I_{0}\right]\boldsymbol{\Psi}= & \,\,0\,\,,\label{eq:31}
\end{align}
which gives
\begin{align}
\Psi_{0}(E_{0},\overset{\rightarrow}{L})= & \,\,\frac{i\lambda\left(\overrightarrow{e}\cdot\overrightarrow{L}\right)}{E_{0}-\lambda^{2}I_{0}}\Psi_{2}(E_{0},\overset{\rightarrow}{L})\,\,,\label{eq:32}\\
\Psi_{1}(E_{0},\overset{\rightarrow}{L})= & \,\,\frac{i\lambda\left(\overrightarrow{e}\cdot\overrightarrow{L}\right)}{E_{0}-\lambda^{2}I_{0}}\Psi_{3}(E_{0},\overset{\rightarrow}{L})\,\,,\label{eq:33}\\
\Psi_{2}(E_{0},\overset{\rightarrow}{L})= & \,\,\frac{i\lambda\left(\overrightarrow{e}\cdot\overrightarrow{L}\right)}{E_{0}+\lambda^{2}I_{0}}\Psi_{0}(E_{0},\overset{\rightarrow}{L})\,\,,\label{eq:34}\\
\Psi_{3}(E_{0},\overset{\rightarrow}{L})= & \,\,\frac{i\lambda\left(\overrightarrow{e}\cdot\overrightarrow{L}\right)}{E_{0}+\lambda^{2}I_{0}}\Psi_{1}(E_{0},\overset{\rightarrow}{L})\,\,.\label{eq:35}
\end{align}
Equations (\ref{eq:32}) and (\ref{eq:33}) give the negative energy
solutions for the anti-fermions while equations (\ref{eq:34}) and
(\ref{eq:35}) give the positive energy solutions for fermions. Accordingly,
by equating vector coefficient ($e_{j}$) in equation (29) the momentum
like solutions for Dirac rotating particles can be expressed as
\begin{align}
\Psi_{0}(E_{0},\overset{\rightarrow}{L})= & \,\,\frac{\left[e_{j}E_{0}+\lambda\left(\overrightarrow{e}\times\overrightarrow{L}\right)_{j}\right]}{i\lambda\left(L_{j}-\lambda I_{j}\right)}\Psi_{2}(E_{0},\overset{\rightarrow}{L})\thinspace\thinspace,\label{eq:36}\\
\Psi_{1}(E_{0},\overset{\rightarrow}{L})= & \,\,\frac{\left[e_{j}E_{0}+\lambda\left(\overrightarrow{e}\times\overrightarrow{L}\right)_{j}\right]}{i\lambda\left(L_{j}-\lambda I_{j}\right)}\Psi_{3}(E_{0},\overset{\rightarrow}{L})\thinspace\thinspace,\label{eq:37}\\
\Psi_{2}(E_{0},\overset{\rightarrow}{L})= & \,\,\frac{\left[e_{j}E_{0}+\lambda\left(\overrightarrow{e}\times\overrightarrow{L}\right)_{j}\right]}{i\lambda\left(L_{j}-\lambda I_{j}\right)}\Psi_{0}(E_{0},\overset{\rightarrow}{L})\thinspace\thinspace,\label{eq:38}\\
\Psi_{3}(E_{0},\overset{\rightarrow}{L})= & \,\,\frac{\left[e_{j}E_{0}+\lambda\left(\overrightarrow{e}\times\overrightarrow{L}\right)_{j}\right]}{i\lambda\left(L_{j}-\lambda I_{j}\right)}\Psi_{1}(E_{0},\overset{\rightarrow}{L})\,\thinspace,\label{eq:39}
\end{align}
where the equations (\ref{eq:36}) and (\ref{eq:37}) represent the
momentum solution in fermionic spinor field while equations (\ref{eq:38})
and (\ref{eq:39}) represent the momentum solution in anti fermionic
spinor field.

\section{Energy-momentum description of rotational Dirac spinors in presence
of electromagnetic field}

Let us start with the quaternionic four-momentum minimum coupling
on the interaction of a charged particle with an electromagnetic (EM)
field as
\begin{align}
P^{\mu}\,\,\rightarrow\,\,\, & \,\thinspace\thinspace P^{\mu}-\frac{Q}{c}V^{\mu}\,\thinspace\equiv\thinspace\Pi^{\mu}\,\,,\thinspace\thinspace\thinspace\,(\mu=0,1,2,3)\,\thinspace,\label{eq:40}
\end{align}
where $P^{\mu}$ is the canonical momentum, $Q$ is the generalized
charge of dyons $\left(Q=\mathfrak{e}-ic\left(\mathfrak{m}\right)\right)$
where $\mathfrak{e}$ is the electric charge and $\mathfrak{m}$ is
the magnetic charge and $\Pi^{\mu}$ is the kinetic momentum. Now,
considering the rotating Dirac fermions, as an analogous, the four
angular momentum can be transferred as $L^{\mu}:\rightarrow\,\,L^{\mu}-\frac{Q}{\lambda}\left(X^{\mu}V^{\mu}\right)\,\thinspace\equiv\thinspace\Pi_{L}^{\mu}$,
where $\Pi_{L}^{\mu}$ is the kinetic angular momentum. Thus we can
express the quaternionic form of rotational four momentum in presence
of EM-field $(L_{q_{em}})$ as
\begin{align}
L_{q_{em}}= & \thinspace\thinspace L_{q}-\mathscr{L}\thinspace\thinspace,\label{eq:41}
\end{align}
where $\mathscr{L}=\frac{Q}{\lambda}\left(X_{q}\circ V_{q}\right)$
is the electromagnetic interaction term coupled with quaternionic
angular momentum, such that 
\begin{align}
\mathscr{L}= & \,\,\frac{Q}{\lambda}\left[\,e_{0}\left(X_{0}\Phi-\overrightarrow{X}\cdot\overrightarrow{V}\right)+e_{j}\left(X_{0}\overrightarrow{V}+\Phi\overrightarrow{X}+\left(\overrightarrow{X}\times\overrightarrow{V}\right)_{j}\right)\right]\,\,.\label{eq:42}
\end{align}
On substituting the values of $V_{q}$ given in (\ref{eq:19}), we
have
\begin{align}
\mathscr{L}= & \frac{Q}{\lambda}\left[e_{0}\left\{ \left(X_{0}\phi_{e}-\overset{\rightarrow}{X}\cdot\overset{\rightarrow}{A}\right)-i\lambda(X_{0}\phi_{m}-\overset{\rightarrow}{X}\cdot\overset{\rightarrow}{B})\right\} \right.\nonumber \\
+ & \left.e_{j}\left\{ X_{0}\overset{\rightarrow}{A}+\phi_{e}\overset{\rightarrow}{X}+\left(\overset{\rightarrow}{X}\times\overset{\rightarrow}{A}\right)_{j}-i\lambda\left(X_{0}\overset{\rightarrow}{B}+\phi_{m}\overset{\rightarrow}{X}+\left(\overset{\rightarrow}{X}\times\overset{\rightarrow}{B}\right)_{j}\right)\right\} \right]\,\,.\label{eq:43}
\end{align}
The quaternionic rotational four-momentum can be obtained in presence
of EM-field, i.e.,
\begin{align}
L_{q_{em}}= & \thinspace\thinspace\thinspace e_{0}\left[\left(X_{0}p_{0}-\overset{\rightarrow}{X}\cdot\overset{\rightarrow}{p}\right)-\frac{Q}{\lambda}\left(X_{0}V_{0}-\overrightarrow{X}\cdot\overrightarrow{V}\right)\right]\nonumber \\
+ & e_{j}\left[\left(X_{0}\overrightarrow{p}+p_{0}\overrightarrow{X}+\left(\overrightarrow{X}\times\overrightarrow{p}\right)_{j}\right)-\frac{Q}{\lambda}\left(X_{0}\overrightarrow{V}+\Phi\overrightarrow{X}+\left(\overrightarrow{X}\times\overrightarrow{V}\right)_{j}\right)\right]\,\thinspace.\label{eq:44}
\end{align}

\noindent Now, we examine at the quaternionic scalar portion known
as rotational EM-energy $(E_{em})$ that corresponds to the basis
element $e_{0}$ as
\begin{align}
E_{em}= & \,\thinspace\left[\left(X_{0}p_{0}-\overset{\rightarrow}{X}\cdot\overset{\rightarrow}{p}\right)-\frac{Q}{\lambda}\left\{ \left(X_{0}\phi_{e}-\overset{\rightarrow}{X}\cdot\overset{\rightarrow}{A}\right)-i\lambda\left(X_{0}\phi_{m}-\overset{\rightarrow}{X}\cdot\overset{\rightarrow}{B}\right)\right\} \right]\nonumber \\
= & \,\,E_{0}-\left(E_{e}-i\lambda E_{m}\right)\:,\label{eq:45}
\end{align}
where the first term $E_{0}$ is the quaternionic rotational energy
given in equation (\ref{eq:24}), the second term indicates the rotational
electric energy $(E_{e})$ and the third term indicates the rotational
magnetic energy $(E_{m})$, so that
\begin{align}
E_{e}= & \,\thinspace\frac{Q}{\lambda}\left[\left(X_{0}\phi_{e}-\overset{\rightarrow}{X}\cdot\overset{\rightarrow}{A}\right)\right]\,\,,\label{eq:46}\\
E_{m}= & \thinspace\thinspace\frac{Q}{\lambda}\left[\left(X_{0}\phi_{m}-\overset{\rightarrow}{X}\cdot\overset{\rightarrow}{B}\right)\right]\,\,.\label{eq:47}
\end{align}

\noindent Consider the scalar functions $\phi_{e}^{X}\mapsto\left(X_{0}\phi_{e}-\overset{\rightarrow}{X}\cdot\overset{\rightarrow}{A}\right)$
and $\phi_{m}^{X}\mapsto\left(X_{0}\phi_{m}-\overset{\rightarrow}{X}\cdot\overset{\rightarrow}{B}\right)$
are, respectively, the generalized quaternionic rotational electric
and magnetic potentials. Then, the equation (\ref{eq:45}) can be
written as $E_{em}=E_{0}-\frac{Q}{\lambda}\Phi^{X}$, where $\Phi^{X}=\phi_{e}^{X}-i\lambda\phi_{m}^{X}$
is the quaternionic rotational generalized scalar potential. Similarly,
in equation (\ref{eq:44}) the quaternionic vector portion denoted
the EM-angular momentum ($\overrightarrow{L}_{em}$) corresponding
to basis element $e_{j}$, i.e.,
\begin{align}
\overrightarrow{L}_{em}= & \thinspace\thinspace\left[X_{0}\overrightarrow{p}+p_{0}\overrightarrow{X}+\left(\overrightarrow{X}\times\overrightarrow{p}\right)-\frac{Q}{\lambda}\left\{ \left(X_{0}\overset{\rightarrow}{A}+\phi_{e}\overset{\rightarrow}{X}+\left(\overset{\rightarrow}{X}\times\overset{\rightarrow}{A}\right)\right)-i\lambda\left(X_{0}\overset{\rightarrow}{B}+\phi_{m}\overset{\rightarrow}{X}+\left(\overset{\rightarrow}{X}\times\overset{\rightarrow}{B}\right)\right)\right\} \right]\nonumber \\
= & \,\,\overrightarrow{L}-\left(\overrightarrow{L}_{e}-i\lambda\overrightarrow{L}_{m}\right)\,\thinspace.\label{eq:48}
\end{align}
In equation (\ref{eq:48}), $L_{e}$ and $L_{m}$ are the electric
and magnetic angular momentum, respectively. As such,
\begin{align}
\overrightarrow{L}_{e}= & \thinspace\,\frac{Q}{\lambda}\overrightarrow{A}^{X}\,\,,\label{eq:49}\\
\overrightarrow{L}_{m}= & \thinspace\thinspace\frac{Q}{\lambda}\overrightarrow{B}^{X}\,\,,\label{eq:50}
\end{align}

\noindent where $\overrightarrow{A}^{X}\mapsto\left(X_{0}\overset{\rightarrow}{A}+\phi_{e}\overset{\rightarrow}{X}+\left(\overset{\rightarrow}{X}\times\overset{\rightarrow}{A}\right)\right)$
can be considered as quaternionic generalized rotational electric
vector potential and $\overrightarrow{B}^{X}\mapsto\left(X_{0}\overset{\rightarrow}{B}+\phi_{m}\overset{\rightarrow}{X}+\left(\overset{\rightarrow}{X}\times\overset{\rightarrow}{B}\right)\right)$
is considered as quaternionic rotational generalized magnetic vector
potential. Thus, the generalized quaternionic angular momentum in
presence of EM-field becomes $\overrightarrow{L}_{em}:\longmapsto\left(\overrightarrow{L}-\frac{Q}{\lambda}\overrightarrow{V}^{X}\right)$
along with vector potential $\overrightarrow{V}^{X}=\overrightarrow{A}^{X}-i\lambda\overrightarrow{B}^{X}$.

\section{Generalized QRD solutions in presence of EM-field}

Let us rewrite the generalized Dirac equation in the presence of an
EM-field associated with four-momentum minimum coupling (\ref{eq:41})
for rotating fermions as
\begin{align}
\left[H_{q}\circ\left(L_{q}-\mathscr{L}\right)-\mathfrak{B}\lambda^{2}I_{q}\right]\circ\boldsymbol{\Psi}= & \,\thinspace0\thinspace\,.\label{eq:51}
\end{align}
It is worth noting that equation (\ref{eq:51}) is rotational analog
of the standard Dirac equation in presence of external EM-field as
$\left(\overrightarrow{\alpha}\cdot\left(\overrightarrow{p}-\frac{Q}{c}\overset{\rightarrow}{V}\right)\right.$$\left.-\mathscr{\beta}mc^{2}\right)\psi=0.$
As a result, using the equations (\ref{eq:44}), the equation (\ref{eq:51})
gives the generalized QRD equation in EM-field,
\begin{align}
\left[e_{0}\left[D^{0}\left(H\right)\left(E_{0}-\frac{Q}{\lambda}\Phi^{X}\right)-\lambda\left(\overrightarrow{D}\left(H\right)\cdot\left(\overrightarrow{L}-\frac{Q}{\lambda}\overrightarrow{V}^{X}\right)\right)-\mathfrak{B}\lambda^{2}I_{0}\right]+e_{j}\left[\lambda D^{0}\left(H\right)\left(\overrightarrow{L}-\frac{Q}{\lambda}\overrightarrow{V}^{X}\right)\right.\right. & +\nonumber \\
\left.D^{j}\left(H\right)\left(E_{0}-\frac{Q}{\lambda}\Phi^{X}\right)+\left.\lambda\left(\overrightarrow{D}\left(H\right)\times\left(\overrightarrow{L}-\frac{Q}{\lambda}\overrightarrow{V}^{X}\right)\right)_{j}-\mathfrak{B}\lambda^{2}I_{j}\right]\right]\circ\boldsymbol{\Psi}=\, & 0\:,\label{eq:52}
\end{align}
where the components of quaternionic electromagnetic coupling term
$H_{q}\circ\left(L_{q}-\mathscr{L}\right)=H_{q}\circ L_{q_{em}}$
can be expressed as 
\begin{align}
H_{q}\circ L_{q_{em}}= & \,\,e_{0}\left[D^{0}\left(H\right)\left(E_{0}-\frac{Q}{\lambda}\Phi^{X}\right)-\lambda D^{1}\left(H\right)\left(L_{1}-\frac{Q}{\lambda}\overrightarrow{V_{1}}^{X}\right)-\lambda D^{2}\left(H\right)\left(L_{2}-\frac{Q}{\lambda}\overrightarrow{V_{2}}^{X}\right)-\lambda D^{3}\left(H\right)\left(L_{3}-\frac{Q}{\lambda}\overrightarrow{V_{3}}^{X}\right)\right]\nonumber \\
+ & e_{1}\left[\lambda D^{0}\left(H\right)\left(L_{1}-\frac{Q}{\lambda}\overrightarrow{V_{1}}^{X}\right)+D^{1}\left(H\right)\left(E_{0}-\frac{Q}{\lambda}\Phi^{X}\right)+\lambda D^{2}\left(H\right)\left(L_{3}-\frac{Q}{\lambda}\overrightarrow{V_{3}}^{X}\right)-\lambda D^{3}\left(H\right)\left(L_{2}-\frac{Q}{\lambda}\overrightarrow{V_{2}}^{X}\right)\right]\nonumber \\
+ & e_{2}\left[\lambda D^{0}\left(H\right)\left(L_{2}-\frac{Q}{\lambda}\overrightarrow{V_{2}}^{X}\right)+D^{2}\left(H\right)\left(E_{0}-\frac{Q}{\lambda}\Phi^{X}\right)+\lambda D^{3}\left(H\right)\left(L_{1}-\frac{Q}{\lambda}\overrightarrow{V_{1}}^{X}\right)-\lambda D^{1}\left(H\right)\left(L_{3}-\frac{Q}{\lambda}\overrightarrow{V_{3}}^{X}\right)\right]\nonumber \\
+ & e_{3}\left[\lambda D^{0}\left(H\right)\left(L_{3}-\frac{Q}{\lambda}\overrightarrow{V_{3}}^{X}\right)+D^{3}\left(H\right)\left(E_{0}-\frac{Q}{\lambda}\Phi^{X}\right)+\lambda D^{1}\left(H\right)\left(L_{2}-\frac{Q}{\lambda}\overrightarrow{V_{2}}^{X}\right)-\lambda D^{2}\left(H\right)\left(L_{1}-\frac{Q}{\lambda}\overrightarrow{V_{1}}^{X}\right)\right]\thinspace,\label{eq:53}
\end{align}

\noindent It is important to note that in the presence of an electromagnetic
field, the generalized QRD equation has a dual structure that includes
both the EM-rotational energy (scalar component) and the EM-rotational
momentum (vector component) of the Dirac fermions. Now, with adding
the two and four-component spinors field, we will construct the energy
and momentum like solutions of generalized QRD equation (\ref{eq:52})
in the following subsections.

\subsection{Energy solutions corresponding to real quaternion}

In this case, we must equate the scalar coefficient of $e_{0}$ as
to calculate the energy solution of the generalized QRD equation (\ref{eq:52})
as
\begin{align}
\left[D^{0}\left(H\right)\left(E_{0}-\frac{Q}{\lambda}\Phi^{X}\right)-\lambda\left(\overrightarrow{D}\left(H\right)\cdot\left(\overrightarrow{L}-\frac{Q}{\lambda}\overrightarrow{V}^{X}\right)\right)-\mathfrak{B}\lambda^{2}I_{0}\right]\boldsymbol{\Psi}= & \,\,0\,\,.\label{eq:54}
\end{align}
On substituting the values of quaternionic $D$-matrices, we obtain

\begin{align}
\left(\begin{array}{cc}
\left[\left(E_{0}-\frac{Q}{\lambda}\Phi^{X}\right)-\lambda^{2}I_{0}\right] & -i\lambda\left[\overrightarrow{e}\cdot\left(\overrightarrow{L}-\frac{Q}{\lambda}\overrightarrow{V}^{X}\right)\right]\\
-i\lambda\left[\overrightarrow{e}\cdot\left(\overrightarrow{L}-\frac{Q}{\lambda}\overrightarrow{V}^{X}\right)\right] & \left[\left(E_{0}-\frac{Q}{\lambda}\Phi^{X}\right)+\lambda^{2}I_{0}\right]
\end{array}\right)\left(\begin{array}{c}
\Psi_{K}\\
\Psi_{L}
\end{array}\right)= & \,\,0\,\,.\label{eq:55}
\end{align}
If we use a four-component quaternionic spinors field, then we get
\begin{align}
\Psi_{0}\left(E_{em},\overrightarrow{L}_{em}\right)= & \,\,\frac{i\lambda\left(\overrightarrow{e}\cdot\left(\overrightarrow{L}-\frac{Q}{\lambda}\overrightarrow{V}^{X}\right)\right)}{\left(E_{0}-\frac{Q}{\lambda}\Phi^{X}\right)-\lambda^{2}I_{0}}\Psi_{2}\left(E_{em},\overrightarrow{L}_{em}\right)\,\,,\label{eq:56}\\
\Psi_{1}\left(E_{em},\overrightarrow{L}_{em}\right)= & \,\,\frac{i\lambda\left(\overrightarrow{e}\cdot\left(\overrightarrow{L}-\frac{Q}{\lambda}\overrightarrow{V}^{X}\right)\right)}{\left(E_{0}-\frac{Q}{\lambda}\Phi^{X}\right)-\lambda^{2}I_{0}}\Psi_{3}\left(E_{em},\overrightarrow{L}_{em}\right)\,\,,\label{eq:57}\\
\Psi_{2}\left(E_{em},\overrightarrow{L}_{em}\right)= & \,\,\frac{i\lambda\left(\overrightarrow{e}\cdot\left(\overrightarrow{L}-\frac{Q}{\lambda}\overrightarrow{V}^{X}\right)\right)}{\left(E_{0}-\frac{Q}{\lambda}\Phi^{X}\right)+\lambda^{2}I_{0}}\Psi_{0}\left(E_{em},\overrightarrow{L}_{em}\right)\,\,,\label{eq:58}\\
\Psi_{3}\left(E_{em},\overrightarrow{L}_{em}\right)= & \,\,\frac{i\lambda\left(\overrightarrow{e}\cdot\left(\overrightarrow{L}-\frac{Q}{\lambda}\overrightarrow{V}^{X}\right)\right)}{\left(E_{0}-\frac{Q}{\lambda}\Phi^{X}\right)+\lambda^{2}I_{0}}\Psi_{1}\left(E_{em},\overrightarrow{L}_{em}\right)\,\,.\label{eq:59}
\end{align}
Now, the entire spinor solutions for the QRD equation may now be calculated
by introducing the plane wavefunction ($\boldsymbol{\psi}$) as $\boldsymbol{\Psi}=\chi_{s}\boldsymbol{\psi}$
where $\chi_{s}$ is the Dirac spinor in presence of EM-field. To
calculate the rotational energy solutions of generalized quaternionic
spinors with spin up state \cite{key-43}, we put $\Psi_{0}=1$, $\Psi_{1}=0$
for positive energy and $\Psi_{2}=1$, $\Psi_{3}=0$ for negative
energy Dirac spinors as
\begin{align}
\boldsymbol{\Psi}:\rightarrow\boldsymbol{\Psi_{+}^{\uparrow}}=\,\,N_{+} & \left(\begin{array}{c}
1\\
0\\
\frac{i\lambda\left[\overrightarrow{e}\cdot\left(\overrightarrow{L}-\frac{Q}{\lambda}\overrightarrow{V}^{X}\right)\right]}{\left(E_{0}-\frac{Q}{\lambda}\Phi^{X}\right)+\lambda^{2}I_{0}}\\
0
\end{array}\right)e^{\frac{i}{\hslash}\left(\overset{\rightarrow}{k}\cdot\overset{\rightarrow}{r}-\omega t\right)}\,\,,\label{eq:60}\\
\boldsymbol{\Psi}:\rightarrow\boldsymbol{\Psi_{-}^{\uparrow}}=\,\,N_{-} & \left(\begin{array}{c}
\frac{i\lambda\left[\overrightarrow{e}\cdot\left(\overrightarrow{L}-\frac{Q}{\lambda}\overrightarrow{V}^{X}\right)\right]}{\left(E_{0}-\frac{Q}{\lambda}\Phi^{X}\right)-\lambda^{2}I_{0}}\\
0\\
1\\
0
\end{array}\right)e^{\frac{i}{\hslash}\left(\overset{\rightarrow}{k}\cdot\overset{\rightarrow}{r}-\omega t\right)}\thinspace\thinspace,\label{eq:61}
\end{align}
where $N_{+}=\frac{\left(E_{0}-\frac{Q}{\lambda}\Phi^{X}\right)+\lambda^{2}I_{0}}{\sqrt{\left[\left(E_{0}-\frac{Q}{\lambda}\Phi^{X}\right)+\lambda^{2}I_{0}\right]^{2}+\lambda^{2}\left(\overrightarrow{L}-\frac{Q}{\lambda}\overrightarrow{V}^{X}\right)^{2}}}$
and $N_{-}=\frac{\left(E_{0}-\frac{Q}{\lambda}\Phi^{X}\right)-\lambda^{2}I_{0}}{\sqrt{\left[\left(E_{0}-\frac{Q}{\lambda}\Phi^{X}\right)-\lambda^{2}I_{0}\right]^{2}+\lambda^{2}\left(\overrightarrow{L}-\frac{Q}{\lambda}\overrightarrow{V}^{X}\right)^{2}}}$
are the normalization constants. Similarly, for spin down state we
put $\Psi_{0}=0$ $\Psi_{1}=1$ for positive energy and $\Psi_{2}=0$
$\Psi_{3}=1$ for negative energy as
\begin{align}
\boldsymbol{\Psi}:\rightarrow\boldsymbol{\Psi_{+}^{\downarrow}}=\,\,N_{+} & \left(\begin{array}{c}
0\\
1\\
0\\
\frac{i\lambda\left[\overrightarrow{e}\cdot\left(\overrightarrow{L}-\frac{Q}{\lambda}\overrightarrow{V}^{X}\right)\right]}{\left(E_{0}-\frac{Q}{\lambda}\Phi^{X}\right)+\lambda^{2}I_{0}}
\end{array}\right)e^{\frac{i}{\hslash}\left(\overset{\rightarrow}{k}\cdot\overset{\rightarrow}{r}-\omega t\right)}\thinspace\thinspace,\label{eq:62}\\
\boldsymbol{\Psi}:\rightarrow\boldsymbol{\Psi_{-}^{\downarrow}}=\,\,N_{-} & \left(\begin{array}{c}
0\\
\frac{i\lambda\left[\overrightarrow{e}\cdot\left(\overrightarrow{L}-\frac{Q}{\lambda}\overrightarrow{V}^{X}\right)\right]}{\left(E_{0}-\frac{Q}{\lambda}\Phi^{X}\right)-\lambda^{2}I_{0}}\\
0\\
1
\end{array}\right)e^{\frac{i}{\hslash}\left(\overset{\rightarrow}{k}\cdot\overset{\rightarrow}{r}-\omega t\right)}\,\thinspace.\label{eq:63}
\end{align}

\subsection{Momentum solutions corresponding to pure quaternion}

In order to calculate the rotational momentum solutions of generalized
QRD equation in EM-field, we equate the coefficient of $e_{j}$ in
equation (\ref{eq:52}) as

\begin{align}
\left[\lambda D^{0}\left(H\right)\left(\overrightarrow{L}-\frac{Q}{\lambda}\overrightarrow{V}^{X}\right)+D^{j}\left(H\right)\left(E_{0}-\frac{Q}{\lambda}\Phi^{X}\right)+\lambda\left(\overrightarrow{D}\left(H\right)\times\left(\overrightarrow{L}-\frac{Q}{\lambda}\overrightarrow{V}^{X}\right)\right)_{j}-\mathfrak{B}\lambda^{2}I_{j}\right]\boldsymbol{\Psi}= & \,\,0\:,\label{eq:64}
\end{align}
which gives
\begin{align}
\left(\begin{array}{cc}
\lambda\left[\left(\overrightarrow{L}-\frac{Q}{\lambda}\overrightarrow{V}^{X}\right)-\lambda I_{j}\right] & i\left[e_{j}\left(E_{0}-\frac{Q}{\lambda}\Phi^{X}\right)+\lambda\left[\overrightarrow{e}\times\left(\overrightarrow{L}-\frac{Q}{\lambda}\overrightarrow{V}^{X}\right)\right]_{j}\right]\\
i\left[e_{j}\left(E_{0}-\frac{Q}{\lambda}\Phi^{X}\right)+\lambda\left[\overrightarrow{e}\times\left(\overrightarrow{L}-\frac{Q}{\lambda}\overrightarrow{V}^{X}\right)\right]_{j}\right] & \lambda\left[\left(\overrightarrow{L}-\frac{Q}{\lambda}\overrightarrow{V}^{X}\right)+\lambda I_{j}\right]
\end{array}\right)\left(\begin{array}{c}
\Psi_{K}\\
\Psi_{L}
\end{array}\right)= & \,\,0\,\,.\label{eq:65}
\end{align}
On substituting the values of $\Psi_{K}$ and $\Psi_{L},$ we obtain
four component spinors fields as 
\begin{align}
\Psi_{0}\left(E_{em},\overrightarrow{L}_{em}\right)= & \,\,\frac{-i\left[e_{j}\left(E_{0}-\frac{Q}{\lambda}\Phi^{X}\right)+\lambda\left(\overrightarrow{e}\times\left(\overrightarrow{L}-\frac{Q}{\lambda}\overrightarrow{V}^{X}\right)\right)_{j}\right]}{\lambda\left(\left(\overrightarrow{L}-\frac{Q}{\lambda}\overrightarrow{V}^{X}\right)-\lambda I_{j}\right)}\Psi_{2}\left(E_{em},\overrightarrow{L}_{em}\right)\,\,,\label{eq:66}\\
\Psi_{1}\left(E_{em},\overrightarrow{L}_{em}\right)= & \,\,\frac{-i\left[e_{j}\left(E_{0}-\frac{Q}{\lambda}\Phi^{X}\right)+\lambda\left(\overrightarrow{e}\times\left(\overrightarrow{L}-\frac{Q}{\lambda}\overrightarrow{V}^{X}\right)\right)_{j}\right]}{\lambda\left(\left(\overrightarrow{L}-\frac{Q}{\lambda}\overrightarrow{V}^{X}\right)-\lambda I_{j}\right)}\Psi_{3}\left(E_{em},\overrightarrow{L}_{em}\right)\,\,,\label{eq:67}\\
\Psi_{2}\left(E_{em},\overrightarrow{L}_{em}\right)= & \,\,\frac{-i\left[e_{j}\left(E_{0}-\frac{Q}{\lambda}\Phi^{X}\right)+\lambda\left(\overrightarrow{e}\times\left(\overrightarrow{L}-\frac{Q}{\lambda}\overrightarrow{V}^{X}\right)\right)_{j}\right]}{\lambda\left(\left(\overrightarrow{L}-\frac{Q}{\lambda}\overrightarrow{V}^{X}\right)+\lambda I_{j}\right)}\Psi_{0}\left(E_{em},\overrightarrow{L}_{em}\right)\,\,,\label{eq:68}
\end{align}
\begin{align}
\Psi_{3}\left(E_{em},\overrightarrow{L}_{em}\right)= & \,\,\frac{-i\left[e_{j}\left(E_{0}-\frac{Q}{\lambda}\Phi^{X}\right)+\lambda\left(\overrightarrow{e}\times\left(\overrightarrow{L}-\frac{Q}{\lambda}\overrightarrow{V}^{X}\right)\right)_{j}\right]}{\lambda\left(\left(\overrightarrow{L}-\frac{Q}{\lambda}\overrightarrow{V}^{X}\right)+\lambda I_{j}\right)}\Psi_{1}\left(E_{em},\overrightarrow{L}_{em}\right)\,\,,\label{eq:69}
\end{align}
where equations (\ref{eq:66}) and (\ref{eq:67}) yield the angular
momentum solution for the anti-particle whereas from equations (\ref{eq:68})
and (\ref{eq:69}) yield the momentum solutions for the particle.
The term $e_{j}E_{em}$ is due to the electromagnetic energy of spin
$\frac{1}{2}$ particles and $\lambda\left(\overrightarrow{L}_{em}-\lambda I_{j}\right),\lambda\left(\overrightarrow{L}_{em}+\lambda I_{j}\right)$
are the energy terms for the anti-particles whereas $\left(\overrightarrow{e}\times\overrightarrow{L}_{em}\right)$
can be interpreted as the directional interaction governed by spin
and the orbital angular momentum. As a result, for angular momentum
solutions with rotating fermions in the spin up state, we put $\Psi_{0}=1$,
$\Psi_{1}=0$ for fermionic angular momentum and $\Psi_{2}=1$, $\Psi_{3}=0$
for anti-fermionic angular momentum as 
\begin{align}
\boldsymbol{\Psi}:\rightarrow\boldsymbol{\Psi_{+}^{\uparrow}}=\,\,M_{+} & \left(\begin{array}{c}
1\\
0\\
\frac{-i\left[e_{j}\left(E_{0}-\frac{Q}{\lambda}\Phi^{X}\right)+\lambda\left(\overrightarrow{e}\times\left(\overrightarrow{L}-\frac{Q}{\lambda}\overrightarrow{V}^{X}\right)\right)_{j}\right]}{\lambda\left(\left(\overrightarrow{L}-\frac{Q}{\lambda}\overrightarrow{V}^{X}\right)+\lambda I_{j}\right)}\\
0
\end{array}\right)e^{\frac{i}{\hslash}\left(\overset{\rightarrow}{k}\cdot\overset{\rightarrow}{r}-\omega t\right)}\thinspace,\label{eq:70}\\
\boldsymbol{\Psi}:\rightarrow\boldsymbol{\Psi_{-}^{\uparrow}}=\,\,M_{-} & \left(\begin{array}{c}
\frac{-i\left[e_{j}\left(E_{0}-\frac{Q}{\lambda}\Phi^{X}\right)+\lambda\left(\overrightarrow{e}\times\left(\overrightarrow{L}-\frac{Q}{\lambda}\overrightarrow{V}^{X}\right)\right)_{j}\right]}{\lambda\left(\left(\overrightarrow{L}-\frac{Q}{\lambda}\overrightarrow{V}^{X}\right)-\lambda I_{j}\right)}\\
0\\
1\\
0
\end{array}\right)e^{\frac{i}{\hslash}\left(\overset{\rightarrow}{k}\cdot\overset{\rightarrow}{r}-\omega t\right)}\,.\label{eq:71}
\end{align}
where the normalization constants are 
\begin{align*}
M_{+}= & \thinspace\thinspace\frac{\lambda\left(\left(\overrightarrow{L}-\frac{Q}{\lambda}\overrightarrow{V}^{X}\right)+\lambda I_{j}\right)}{\sqrt{\left[\lambda\left(\left(\overrightarrow{L}-\frac{Q}{\lambda}\overrightarrow{V}^{X}\right)+\lambda I_{j}\right)\right]^{2}-\left[e_{j}\left(E_{0}-\frac{Q}{\lambda}\Phi^{X}\right)+\lambda\left(\overrightarrow{e}\times\left(\overrightarrow{L}-\frac{Q}{\lambda}\overrightarrow{V}^{X}\right)\right)_{j}\right]^{2}}}\thinspace\thinspace,\\
M_{-}= & \,\,\frac{\lambda\left(\left(\overrightarrow{L}-\frac{Q}{\lambda}\overrightarrow{V}^{X}\right)-\lambda I_{j}\right)}{\sqrt{\left[\lambda\left(\left(\overrightarrow{L}-\frac{Q}{\lambda}\overrightarrow{V}^{X}\right)-\lambda I_{j}\right)\right]^{2}-\left[e_{j}\left(E_{0}-\frac{Q}{\lambda}\Phi^{X}\right)+\lambda\left(\overrightarrow{e}\times\left(\overrightarrow{L}-\frac{Q}{\lambda}\overrightarrow{V}^{X}\right)\right)_{j}\right]^{2}}}\thinspace\thinspace.
\end{align*}
Correspondingly, to write the angular momentum solutions of rotating
fermions in the spin down state, we put $\Psi_{0}=0$, $\Psi_{1}=1$
for fermionic angular momentum and $\Psi_{2}=0$, $\Psi_{3}=1$ for
anti-fermionic angular momentum as
\begin{align}
\boldsymbol{\Psi}:\rightarrow\boldsymbol{\Psi_{+}^{\downarrow}}=\,\,M_{+} & \left(\begin{array}{c}
0\\
1\\
0\\
\frac{-i\left[e_{j}\left(E_{0}-\frac{Q}{\lambda}\Phi^{X}\right)+\lambda\left[\overrightarrow{e}\times\left(\overrightarrow{L}-\frac{Q}{\lambda}\overrightarrow{V}^{X}\right)\right]_{j}\right]}{\lambda\left(\left(\overrightarrow{L}-\frac{Q}{\lambda}\overrightarrow{V}^{X}\right)+\lambda I_{j}\right)}
\end{array}\right)e^{\frac{i}{\hslash}\left(\overset{\rightarrow}{k}\cdot\overset{\rightarrow}{r}-\omega t\right)}\thinspace,\label{eq:72}
\end{align}
\begin{align}
\boldsymbol{\Psi}:\rightarrow\boldsymbol{\Psi_{-}^{\downarrow}}=\,\,M_{-} & \left(\begin{array}{c}
0\\
\frac{-i\left[e_{j}\left(E_{0}-\frac{Q}{\lambda}\Phi^{X}\right)+\lambda\left[\overrightarrow{e}\times\left(\overrightarrow{L}-\frac{Q}{\lambda}\overrightarrow{V}^{X}\right)\right]_{j}\right]}{\lambda\left(\left(\overrightarrow{L}-\frac{Q}{\lambda}\overrightarrow{V}^{X}\right)-\lambda I_{j}\right)}\\
0\\
1
\end{array}\right)e^{\frac{i}{\hslash}\left(\overset{\rightarrow}{k}\cdot\overset{\rightarrow}{r}-\omega t\right)}\,\,.\label{eq:73}
\end{align}
Now, using fundamental symmetries and conservation rules, we will
examine the validity of the foregoing QRD equations and their solutions
in the EM field.

\subsubsection{Lorentz-Poincaré Invariance}

The Lorentz transformation can be expressed using a $4\times4$ transformation
matrix $\varLambda$ as
\begin{align}
\left(X^{\mu}\right)^{\prime}= & \sum_{\nu=0}^{3}\left(\Lambda_{\nu}^{\mu}\right)X^{\nu}\,\,,\thinspace\thinspace\,(\mu,\nu=0,1,2,3),\label{eq:74}
\end{align}
where
\begin{align}
\varLambda_{\nu}^{\mu}= & \left(\begin{array}{cccc}
\cosh\omega & -i\sinh\omega & 0 & 0\\
i\sinh\omega & \cosh\omega & 0 & 0\\
0 & 0 & 1 & 0\\
0 & 0 & 0 & 1
\end{array}\right)\,\thinspace.\label{eq:75}
\end{align}
Here $\omega$ denotes the boost parameter. To check the Lorentz covariance,
we can use the covariant form of the QRD equation as follows:
\begin{align}
\left[\gamma^{\mu}\left(L_{q}^{\mu}-\frac{Q}{\lambda}(X_{q}^{\mu}V_{q}^{\mu})\right)-\lambda I_{q}\right]\boldsymbol{\Psi}= & \,\,0\thinspace\thinspace.\label{eq:76}
\end{align}
Here $\gamma^{\mu}$ are the Dirac gamma matrices which can be represented
in terms of quaternionic D-matrices as
\begin{align}
\gamma^{0}= & \thinspace\mathfrak{B}=\thinspace\left(\begin{array}{cc}
1 & 0\\
0 & -1
\end{array}\right)\,\thinspace\mbox{and }\gamma^{j}=\thinspace\mathfrak{B}D^{j}(H)=\thinspace\left(\begin{array}{cc}
0 & ie_{j}\\
-ie_{j} & 0
\end{array}\right)\,.\label{eq:77}
\end{align}
Now, we have
\begin{align}
\left[\gamma^{\mu}\left(L_{q}^{\mu^{\prime}}-\frac{Q}{\lambda}(X_{q}^{\mu^{\prime}}V_{q}^{\mu^{\prime}})\right)-\lambda I_{q}\right]\boldsymbol{\Psi^{\prime}}= & \thinspace\thinspace0\thinspace\thinspace,\label{eq:78}
\end{align}
with
\begin{align}
\boldsymbol{\Psi^{\prime}=} & \thinspace\thinspace\boldsymbol{S\,\Psi}\:,\label{eq:79}
\end{align}
where $\boldsymbol{S}$ is a $4\times4$ matrix. Therefore, we may
write
\begin{align}
\boldsymbol{S^{-1}}\left[\gamma^{\mu}\boldsymbol{S}\varLambda_{\nu}^{\mu}\left(L_{q}^{\mu}-\frac{Q}{\lambda}\left(X_{q}^{\mu}V_{q}^{\mu}\right)\right)-\lambda I_{q}\boldsymbol{S}\right]\boldsymbol{\Psi}= & \,\,0\thinspace\thinspace,\label{eq:80}
\end{align}
which shows
\begin{align}
\boldsymbol{S^{-1}}\gamma^{\mu}\boldsymbol{S}\left(L_{q}^{\mu}-\frac{Q}{\lambda}\left(X_{q}^{\mu}V_{q}^{\mu}\right)\right)= & \,\thinspace\varLambda_{\nu}^{\mu}\gamma^{\nu}\left(L_{q}^{\mu}-\frac{Q}{\lambda}\left(X_{q}^{\mu}V_{q}^{\mu}\right)\right)\thinspace\thinspace.\label{eq:81}
\end{align}
where the corresponding term $(\lambda I_{q})$ is treated as a constant
during the transformation in order to keep the covariant form of the
Dirac equation. Furthermore, the Lorentz boost can define $\boldsymbol{S}$
for a Lorentz transformation along the $e_{1}$-axis as
\begin{align}
\boldsymbol{S}= & \,\,\cosh\frac{\omega}{2}+iD^{1}(H)\sinh\frac{\omega}{2}\,\thinspace,\label{eq:82}
\end{align}
where $D^{1}(H)\sim\left(\begin{array}{cc}
0 & ie_{1}\\
ie_{1} & 0
\end{array}\right)$. As a result, under Lorentz transformation, the generalized QRD equation
in the EM-field is well invariant.

\subsubsection{Gauge Invariance}

Let us consider the analogous gauge transformation of fermionic spinor
field as
\begin{align}
\boldsymbol{\Psi^{\prime}=} & \thinspace\thinspace\exp\left(-i\frac{Q}{\hslash\lambda}\xi(\overset{\rightarrow}{r},t)\right)\boldsymbol{\Psi}\,\,,\label{eq:83}
\end{align}
where $\xi(\overset{\rightarrow}{r},t)$ is an arbitrary differentiable
gauge function in Minkowski space-time. The gauge transformation of
the two four-potentials is also introduced as
\begin{align}
\overrightarrow{A}^{\prime}= & \thinspace\thinspace\overset{\rightarrow}{A}-\nabla\xi\thinspace\thinspace\thinspace\mbox{and}\thinspace\thinspace\thinspace\phi_{e}^{\prime}=\,\,\phi_{e}+\left(\frac{\partial\xi}{\partial t}\right)\,\thinspace,\label{eq:84}\\
\overrightarrow{B}^{\prime}= & \thinspace\,\overset{\rightarrow}{B}-\nabla\xi\thinspace\thinspace\thinspace\mbox{and}\thinspace\thinspace\thinspace\phi_{m}^{\prime}=\,\,\phi_{m}+\left(\frac{\partial\xi}{\partial t}\right)\,\thinspace,\label{eq:85}
\end{align}
along with

\begin{align}
\left[H_{q}\circ\left(L_{q}-\frac{Q}{\lambda}(X_{q}\circ V_{q}^{\prime})\right)-\mathfrak{B}\lambda^{2}I_{q}\right]\circ\boldsymbol{\Psi^{\prime}}=\,\, & 0\thinspace\,,\label{eq:86}
\end{align}
where $V_{q}^{\prime}$ is the generalized two four-potential transforming
under the gauge transformation and can be defined as 
\begin{align}
\overrightarrow{V}^{\prime}= & \overset{\rightarrow}{V}-\nabla\xi\thinspace\thinspace\thinspace\mbox{and}\thinspace\thinspace\thinspace\Phi^{\prime}=\Phi+\left(\frac{\partial\xi}{\partial t}\right)\,\thinspace,\label{eq:87}
\end{align}
where $\overset{\rightarrow}{V}$ and $\Phi$ are already defined
in equations (\ref{eq:18}) and (\ref{eq:19}).

\subsubsection{Duality Invariance}

To check the duality invariance of generalized QRD equation in EM-field,
let us define the transformation of two field tensors $F_{\mu\nu}$
and $\mathscr{\mathcal{F_{\mu\nu}}}$ as
\begin{align}
F_{\mu\nu}^{\prime}\longrightarrow & \,\,F_{\mu\nu}\cos\alpha-\mathcal{F}_{\mu\nu}\sin\alpha\thinspace\,\label{eq:88}\\
\mathcal{F_{\mu\nu}^{\prime}}\longrightarrow & \thinspace\thinspace\mathcal{F_{\mu\nu}}\sin\alpha+F_{\mu\nu}\cos\alpha\thinspace\thinspace,\thinspace\thinspace\thinspace\left(0\leq\alpha\leq\frac{\pi}{2}\right)\thinspace\thinspace.\label{eq:89}
\end{align}
The electric and magnetic field vectors can be transformed by replacing
$F_{\mu\nu}\longrightarrow\overrightarrow{\mathsf{E}}$ and $\mathcal{F_{\mu\nu}}\longrightarrow\overrightarrow{\mathsf{H}}$
where $\overrightarrow{\mathsf{E}}$ and $\overrightarrow{\mathsf{H}}$
are the electric and magnetic field vectors. Then, the transformation
becomes
\begin{align}
\begin{pmatrix}\overrightarrow{\mathsf{E}}\\
\overrightarrow{\mathsf{H}}
\end{pmatrix}:= & \,\thinspace\begin{pmatrix}\begin{array}{cc}
0 & -i\lambda\\
\frac{i}{\lambda} & 0
\end{array}\end{pmatrix}\begin{pmatrix}\overrightarrow{\mathsf{E}}\\
\overrightarrow{\mathsf{H}}
\end{pmatrix}\thinspace\thinspace.\label{eq:90}
\end{align}
Here we took the general case of $\alpha=\frac{\pi}{2}$. The matrix
$\begin{pmatrix}\begin{array}{cc}
0 & -i\lambda\\
\frac{i}{\lambda} & 0
\end{array}\end{pmatrix}$ is the duality transformation matrix. Accordingly, dual scalar potentials,
dual vector potentials and dual charge can also transform as
\begin{align}
\begin{pmatrix}\phi_{e}\\
\phi_{m}
\end{pmatrix}:= & \thinspace\thinspace\begin{pmatrix}\begin{array}{cc}
0 & -i\lambda\\
\frac{i}{\lambda} & 0
\end{array}\end{pmatrix}\begin{pmatrix}\phi_{e}\\
\phi_{m}
\end{pmatrix}\thinspace\thinspace.\label{eq:91}\\
\begin{pmatrix}\overrightarrow{A}\\
\overrightarrow{B}
\end{pmatrix}:= & \thinspace\thinspace\begin{pmatrix}\begin{array}{cc}
0 & -i\lambda\\
\frac{i}{\lambda} & 0
\end{array}\end{pmatrix}\begin{pmatrix}\overrightarrow{A}\\
\overrightarrow{B}
\end{pmatrix}\thinspace\thinspace,\label{eq:92}\\
\begin{pmatrix}\mathfrak{e}\\
\mathfrak{m}
\end{pmatrix}:= & \,\thinspace\begin{pmatrix}\begin{array}{cc}
0 & -i\lambda\\
\frac{i}{\lambda} & 0
\end{array}\end{pmatrix}\begin{pmatrix}\mathfrak{e}\\
\mathfrak{m}
\end{pmatrix}\,\,.\label{eq:93}
\end{align}
The generalized QRD equation in the presence of an EM-field is thus
well invariant under duality transformation when applying the above
transformation equations.

\subsubsection{CPT Invariance}

In order to check the CPT invariance of the generalized QRD equation
in the EM-field, Table 2 summarizes the transformations of various
physical parameters under parity, time reversal, and charge conjugation
\cite{key-44,key-45}.\textbf{}
\begin{table}
\begin{centering}
\begin{tabular}{cccc}
\hline 
\textbf{Physical quantities} & \textbf{$\mathsf{P}$-symmetry} & \textbf{$\mathsf{C}$-symmetry} & \textbf{$\mathsf{T}$-symmetry}\tabularnewline
\hline 
\hline 
$\overrightarrow{\nabla}$ & $-\overrightarrow{\nabla}$ & $\overrightarrow{\nabla}$ & $\overrightarrow{\nabla}$\tabularnewline
$\partial_{t}$ & $\partial_{t}$ & $\partial_{t}$ & $-\partial_{t}$\tabularnewline
$\lambda$ & $-\lambda$ & $\lambda$ & $-\lambda$\tabularnewline
$X_{0}$ & $X_{0}$ & $X_{0}$ & $-X_{0}$\tabularnewline
$\overset{\rightarrow}{X}$ & $-\overset{\rightarrow}{X}$ & $\overset{\rightarrow}{X}$ & $\overset{\rightarrow}{X}$\tabularnewline
$p_{0}$ & $p_{0}$ & $p_{0}$ & $p_{0}$\tabularnewline
$\overset{\rightarrow}{p}$ & $-\overset{\rightarrow}{p}$ & $\overset{\rightarrow}{p}$ & $-\overset{\rightarrow}{p}$\tabularnewline
$\phi_{e}$ & $\phi_{e}$ & $-\phi_{e}$ & $\phi_{e}$\tabularnewline
$\phi_{m}$ & $-\phi_{m}$ & $-\phi_{m}$ & $-\phi_{m}$\tabularnewline
$\overset{\rightarrow}{A}$ & $-\overset{\rightarrow}{A}$ & $-\overset{\rightarrow}{A}$ & $-\overset{\rightarrow}{A}$\tabularnewline
$\overset{\rightarrow}{B}$ & $\overset{\rightarrow}{B}$ & $-\overset{\rightarrow}{B}$ & $\overset{\rightarrow}{B}$\tabularnewline
$\overrightarrow{\mathsf{E}}$ & $\mathsf{-\overrightarrow{\mathsf{E}}}$ & $-\mathsf{\overrightarrow{\mathsf{E}}}$ & $\mathsf{\overrightarrow{\mathsf{E}}}$\tabularnewline
$\overrightarrow{\mathsf{H}}$ & $\mathsf{\overrightarrow{\mathsf{H}}}$ & $-\overrightarrow{\mathsf{H}}$ & $-\overrightarrow{\mathsf{H}}$\tabularnewline
\hline 
\end{tabular}
\par\end{centering}
\textbf{\caption{\textbf{CPT transformation of various quaternionic physical quantities}}
}
\end{table}
 Apart from the physical quantities, the transformation of the fermionic
spinor fields can be represented as
\begin{align}
\mathsf{P}\boldsymbol{\Psi}(\overset{\rightarrow}{r},t)\mathsf{P^{-1}}= & \thinspace\,\gamma_{0}\boldsymbol{\Psi}(-\overset{\rightarrow}{r},t)\,\thinspace,\label{eq:94}\\
\mathsf{C}\boldsymbol{\Psi}(\overset{\rightarrow}{r},t)\mathsf{C^{-1}}= & \thinspace\thinspace i\gamma_{2}\boldsymbol{\Psi}^{*}(\overset{\rightarrow}{r},t)\,\thinspace,\label{eq:95}\\
\mathsf{T}\boldsymbol{\Psi}(\overset{\rightarrow}{r},t)\mathsf{T^{-1}}= & \thinspace\thinspace i\gamma_{1}\gamma_{3}\boldsymbol{\Psi}(\overset{\rightarrow}{r},-t)\,\,,\label{eq:96}
\end{align}
where $\mathsf{P,C}$ and $\mathsf{T}$ represent parity, charge conjugation
and time reversal operators while $\gamma_{0}=\mathfrak{B},\gamma_{j}=\mathfrak{B}D^{j}(H)$
are equivalent to the quaternionic matrices. From equations (\ref{eq:94})
to (\ref{eq:96}), we get
\begin{align}
\mathsf{CPT}\left[\left(H_{q}\circ L_{q_{em}}-\mathfrak{B}\lambda^{2}I_{q}\right)\circ\boldsymbol{\Psi}\right]\mathsf{T^{-1}P^{-1}C^{-1}}= & \,\thinspace0\thinspace\thinspace,\label{eq:97}
\end{align}
Therefore, under CPT transformation, (\ref{eq:51}) yields an invariant
QRD equation in the EM-field.

\section{Generalized Quaternionic Electromagnetic Moment }

\subsection{Corresponding to the real quaternion}

The Dirac fermions has an electromagnetic moment or electromagnetic
energy that may be determined using the two component form of Pauli's
spinors. To find the electromagnetic moment corresponding to the scalar
coefficient in the generalized QRD equation, we may start with the
given equation (\ref{eq:55}) with condition $E_{em}=E_{em}^{\prime}+\lambda^{2}I_{0}$
where $E_{em}$ is rotational EM-energy, $E_{em}^{\prime}$ is the
rotational kinetic energy and $\lambda^{2}I_{0}$ is rotational analogous
of rest mass energy in EM-field, then one can write
\begin{align}
\left(E_{em}^{\prime}+2\lambda^{2}I_{0}\right)\Psi_{L}-i\lambda\left(\overrightarrow{e}\cdot\overrightarrow{L}_{em}\right)\Psi_{K}= & \,\,0\,\,,\label{eq:98}
\end{align}
Since, in the non-relativistic limit $E_{em}^{\prime}<<\lambda^{2}I_{0},$
we have
\begin{align}
\Psi_{L}\,\,\simeq & \thinspace\,\left(\frac{i\left(\overrightarrow{e}\cdot\overrightarrow{L}_{em}\right)}{2\lambda I_{0}}\right)\Psi_{K}\,\,.\label{eq:99}
\end{align}
As a result, by substituting equation (\ref{eq:99}) in second component
solution of equation (\ref{eq:55}), we obtain
\begin{align}
\left(E_{em}-\lambda^{2}I_{0}\right)\Psi_{K}+ & .\left(\frac{1}{2I_{0}}\left(L_{em}\right)^{2}-\frac{Q\hslash}{2I_{0}\lambda}\left(\overset{\rightarrow}{e}\cdot\overrightarrow{\mathsf{H}}\right)-\frac{iQ\hslash}{2I_{0}}\left(\overset{\rightarrow}{e}\cdot\overrightarrow{\mathsf{E}}\right)\right)\Psi_{K}=0\,\thinspace.\label{eq:100}
\end{align}
This equation can be referred as the generalized Schrodinger-Pauli
like equation where the unperturbed energy term is $\left|\frac{1}{2I_{0}}\left(L_{em}\right)^{2}+\lambda^{2}I_{0}\right|$,
while the perturbed energy terms are $\left|\frac{Q\hslash}{2I_{0}\lambda}\left(\overset{\rightarrow}{e}\cdot\overrightarrow{\mathsf{H}}\right)\right|$
and $\left|\frac{iQ\hslash}{2I_{0}}\left(\overset{\rightarrow}{e}\cdot\overrightarrow{\mathsf{E}}\right)\right|$
corresponding to magnetic and electric dipole moments $\frac{Q\hslash\overset{\rightarrow}{e}}{2I_{0}\lambda}$
and $\frac{iQ\hslash\overset{\rightarrow}{e}}{2I_{0}}$, respectively.
Furthermore, we have found dual perturbed energies, respectively,
the generalized electric dipole energy \cite{key-46} i.e. $\left|\frac{\kappa}{\lambda}\left(\overset{\rightarrow}{e}\cdot\overrightarrow{\mathsf{H}}\right)\right|$
and the generalized magnetic dipole energy i.e. $\left|i\kappa\left(\overset{\rightarrow}{e}\cdot\overrightarrow{\mathsf{E}}\right)\right|$
due to EM-interaction in fermionic field, where $\kappa=\frac{Q\hslash}{2I_{0}}$
is a constant.

\subsection{Corresponding to the pure quaternion}

In this case, we choose vector coefficient in the generalized QRD
equation and applying the condition $\overrightarrow{L}_{em}=\overrightarrow{L}_{em}^{\prime}+\lambda I_{j}$,
and obtain

\begin{align}
\lambda\left(L_{j_{em}}^{\prime}+2\lambda I_{j}\right)\Psi_{L}+i\left[e_{j}E_{em}+\lambda\left(\overrightarrow{e}\times\overrightarrow{L}_{em}\right)_{j}\right]\Psi_{K}= & \,\,0\,\,,\label{eq:101}
\end{align}
For the non-relativistic limit $\lambda L_{j_{em}}^{\prime}<<2\lambda^{2}I_{j},$
then equation (\ref{eq:101}) gives

\begin{align}
\Psi_{L}\approx\thinspace\, & \frac{-i\left[e_{j}E_{em}+\left(\overrightarrow{e}\times\overrightarrow{L}_{em}\right)_{j}\right]}{2\lambda^{2}I_{j}}\Psi_{K}\thinspace\thinspace.\label{eq:102}
\end{align}
Substituting equation (\ref{eq:101}) in second component solution
of equation (\ref{eq:65}), and obtain
\begin{align}
\left[\lambda\overrightarrow{L}_{em}^{\prime}-\frac{1}{2\lambda^{2}I_{j}}\left(E_{em}^{2}\right)+\frac{1}{2I_{j}}\left\{ -2\left(\overrightarrow{L}_{em}^{2}\right)+\frac{Q\hslash}{2I_{j}\lambda}\left(\overset{\rightarrow}{e}\cdot\overrightarrow{\mathsf{H}}\right)+\frac{Q\hslash i}{2I_{j}}\left(\overset{\rightarrow}{e}\cdot\overrightarrow{\mathsf{E}}\right)\right\} +\frac{2e_{j}E_{em}\left(\overrightarrow{e}\times\overrightarrow{L}_{em}\right)_{j}}{2\lambda I_{j}}\right]\Psi_{K}= & \,\,0\,\,.\label{eq:103}
\end{align}
As equation (\ref{eq:100}), the equation (\ref{eq:103}) can describe
the electromagnetic energy in quaternionic rotational momentum space.
The last term denotes the energy corresponding to the directional
spin orbit coupling in the quaternionic angular momentum-space. 

On the other hand, if we consider `rotating dyons' in the generalized
fermionic spinor dual-fields that includes both an electric field
due to electrons and a magnetic field due to magnetic monopoles, then
the following field constituents for generalized Dirac spinors may
be used for unified dyonic fields \cite{key-14,key-35,key-36}:
\begin{align*}
Q_{\text{Dyon}}\,= & \left(\mathfrak{e}-i\lambda\mathfrak{m}\right)\,\,,\,\,\,\,\,\,\,\,\,\,\,\,\,\,\,\,\,\,\,\,(\text{Dyonic Charge})\\
\Phi_{\text{Dyon}}^{X}\,= & \left(\phi_{e}^{X}-i\lambda\phi_{m}^{X}\right)\,\,,\,\,\,\,\,\,\,\,\,\,\,\,\,(\text{Dyonic Scalar Potential})\\
\overrightarrow{V}_{\text{Dyon}}^{X}\,= & \left(\overrightarrow{A}^{X}-i\lambda\overrightarrow{B}^{X}\right)\,\,,\,\,\,\,\,\,\,\,\,\,(\text{Dyonic Vector Potential})\\
E_{\text{Dyon}}\,= & \,\left(E_{e}-i\lambda E_{m}\right)\,\,,\,\,\,\,\,\,\,\,\,\,\,\,\,\,\,(\text{Dyonic Rotational EM-Energy})\\
\overrightarrow{L}_{\text{Dyon}}\,= & \,\left(\overrightarrow{L}_{e}-i\lambda\overrightarrow{L}_{m}\right)\,\,,\,\,\,\,\,\,\,\,\,\,(\text{Dyonic EM-Angular Momentum})\\
E_{\text{Dyon}}^{\text{di}}\,= & \frac{Q\hslash}{2iI_{0}}\left(\overset{\rightarrow}{e}\cdot\overrightarrow{\mathsf{\Psi}}_{\text{Dyon}}\right)\,\,,\,\,\,\,\,\,\,(\text{Dyonic Dipole Energy})\\
\overrightarrow{\mathsf{\Psi}}_{\text{Dyon}}\,= & \left(\overrightarrow{\mathsf{E}}-\frac{i}{\lambda}\overrightarrow{\mathsf{H}}\right)\,\,,\,\,\,\,\,\,\,\,\,\,\,\,\,\,\,\,(\text{Dyonic EM-Field})
\end{align*}
Hence, we examined a symmetrical form of the generalized QRD equation
in the presence of an external EM-field, in which two four-potentials
are recognized as the gauge potentials associated with a particle
(or antiparticle) containing the simultaneous existence of electric
and magnetic charge (monopole) of dyons. 
\begin{sidewaystable}[H]
\begin{centering}
\begin{tabular}{ccc}
\hline 
\textbf{Physical quantities} & \textbf{General Case} & \textbf{Quaternionic-Rotational Case}\tabularnewline
\hline 
\hline 
Dirac equation without EM-field: & $\left(\overrightarrow{\alpha}.\overrightarrow{p}-\beta mc^{2}\right)\psi=0$ & $\left(H_{q}\circ L_{q}-\mathscr{\mathfrak{B}}\lambda^{2}I_{q}\right)\circ\mathtt{\boldsymbol{\Psi}}=0$\tabularnewline
Minimal Substitution: & $\overset{\rightarrow}{p}\rightarrow\overset{\rightarrow}{p}-\frac{e}{c}\overset{\rightarrow}{A}\thinspace\,,E\rightarrow E-e\phi$ & $L^{\mu}:\rightarrow L^{\mu}-\frac{Q}{\lambda}\left(X^{\mu}\circ V^{\mu}\right)\,\thinspace\equiv\thinspace\Pi_{L}^{\mu}\thinspace\thinspace(\mu=0,1,2,3)$\tabularnewline
Dirac equation with EM-field: & $\left(\overrightarrow{\alpha}\cdot\left(\overrightarrow{p}-\frac{e}{c}\overset{\rightarrow}{A}\right)-\mathscr{\beta}mc^{2}\right)\psi=0$ & $\left(H_{q}\circ\left(L_{q}-\frac{Q}{\lambda}(X_{q}\circ V_{q})\right)-\mathfrak{B}\lambda^{2}I_{q}\right)\circ\boldsymbol{\Psi}=0$\tabularnewline
Spinor solutions: & $\psi_{+}^{\uparrow}=\,\,N_{+}\left(\begin{array}{c}
1\\
0\\
\frac{c(\overrightarrow{\sigma}.\overset{\rightarrow}{P})}{E+mc^{2}}\\
0
\end{array}\right),\psi^{\downarrow+}=\,\,N_{+}\left(\begin{array}{c}
0\\
1\\
0\\
\frac{c\left(\overrightarrow{\sigma}\cdot\overrightarrow{P}\right)}{E+mc^{2}}
\end{array}\right)$ & $\boldsymbol{\Psi}_{+}^{\uparrow}=\,\,N_{+}\left(\begin{array}{c}
1\\
0\\
\frac{i\lambda\left(\overrightarrow{e}\cdot\overrightarrow{L}\right)}{E_{0}+\lambda^{2}I_{0}}\\
0
\end{array}\right)\boldsymbol{\Psi}^{\downarrow+}=\,\,N_{+}\left(\begin{array}{c}
0\\
1\\
0\\
\frac{i\lambda\left(\overrightarrow{e}\cdot\overrightarrow{L}\right)}{E_{0}+\lambda^{2}I_{0}}
\end{array}\right)$\tabularnewline
 & $\psi^{\uparrow-}=\,\,N_{-}\left(\begin{array}{c}
\frac{c\left(\overrightarrow{\sigma}\cdot\overrightarrow{P}\right)}{E-mc^{2}}\\
0\\
1\\
0
\end{array}\right),\psi^{\downarrow-}=\,\,N_{-}\left(\begin{array}{c}
0\\
\frac{c\left(\overrightarrow{\sigma}\cdot\overrightarrow{P}\right)}{E-mc^{2}}\\
0\\
1
\end{array}\right)$ & $\boldsymbol{\Psi}^{\uparrow-}=\,\,N_{-}\left(\begin{array}{c}
\frac{i\lambda\left(\overrightarrow{e}\cdot\overrightarrow{L}\right)}{E_{0}-\lambda^{2}I_{0}}\\
0\\
1\\
0
\end{array}\right),\boldsymbol{\Psi}^{\downarrow-}=\,\,N_{-}\left(\begin{array}{c}
0\\
\frac{i\lambda\left(\overrightarrow{e}\cdot\overrightarrow{L}\right)}{E_{0}-\lambda^{2}I_{0}}\\
0\\
1
\end{array}\right)$\tabularnewline
Spinor solutions with EM-field: & $\psi{}_{+}^{\uparrow}=\,\,N_{+}\left(\begin{array}{c}
1\\
0\\
\frac{c\left[\overrightarrow{\sigma}\cdot\left(\overrightarrow{P}-\frac{\mathfrak{e}}{c}\overrightarrow{A}\right)\right]}{\left(E-\mathfrak{e}\phi\right)+mc^{2}}\\
0
\end{array}\right)$ & $\boldsymbol{\Psi_{+}^{\uparrow}}=\,\,N_{+}\left(\begin{array}{c}
1\\
0\\
\frac{i\lambda\left[\overrightarrow{e}\cdot\left(\overrightarrow{L}-\frac{Q}{\lambda}\overrightarrow{V}^{X}\right)\right]}{\left(E_{0}-\frac{Q}{\lambda}\Phi^{X}\right)+\lambda^{2}I_{0}}\\
0
\end{array}\right)$\tabularnewline
 & $\psi{}_{+}^{\downarrow}=\,\,N_{+}\left(\begin{array}{c}
0\\
1\\
0\\
\frac{\lambda\left[\overrightarrow{\sigma}\cdot\left(\overrightarrow{P}-\frac{\mathfrak{e}}{c}\overrightarrow{A}\right)\right]}{\left(E-\mathfrak{e}\phi\right)+mc^{2}}
\end{array}\right)$ & $\boldsymbol{\Psi_{+}^{\downarrow}}=\,\,N_{+}\left(\begin{array}{c}
0\\
1\\
0\\
\frac{i\lambda\left[\overrightarrow{e}\cdot\left(\overrightarrow{L}-\frac{Q}{\lambda}\overrightarrow{V}^{X}\right)\right]}{\left(E_{0}-\frac{Q}{\lambda}\Phi^{X}\right)+\lambda^{2}I_{0}}
\end{array}\right)$\tabularnewline
 & $\psi{}_{-}^{\uparrow}=\,\,N_{-}\left(\begin{array}{c}
\frac{c\left[\overrightarrow{\sigma}\cdot\left(\overrightarrow{P}-\frac{\mathfrak{e}}{c}\overrightarrow{A}\right)\right]}{\left(E-\mathfrak{e}\phi\right)-mc^{2}}\\
0\\
1\\
0
\end{array}\right)$ & $\boldsymbol{\Psi_{-}^{\uparrow}}=\,\,N_{-}\left(\begin{array}{c}
\frac{i\lambda\left[\overrightarrow{e}\cdot\left(\overrightarrow{L}-\frac{Q}{\lambda}\overrightarrow{V}^{X}\right)\right]}{\left(E_{0}-\frac{Q}{\lambda}\Phi^{X}\right)-\lambda^{2}I_{0}}\\
0\\
1\\
0
\end{array}\right)$\tabularnewline
 & $\psi{}_{-}^{\downarrow}=\,\,N_{-}\left(\begin{array}{c}
0\\
\frac{c\left[\overrightarrow{\sigma}\cdot\left(\overrightarrow{P}-\frac{\mathfrak{e}}{c}\overrightarrow{A}\right)\right]}{\left(E-\mathfrak{e}\phi\right)-mc^{2}}\\
0\\
1
\end{array}\right)$ & $\boldsymbol{\Psi_{-}^{\downarrow}}=\,\,N_{-}\left(\begin{array}{c}
0\\
\frac{i\lambda\left[\overrightarrow{e}\cdot\left(\overrightarrow{L}-\frac{Q}{\lambda}\overrightarrow{V}^{X}\right)\right]}{\left(E_{0}-\frac{Q}{\lambda}\Phi^{X}\right)-\lambda^{2}I_{0}}\\
0\\
1
\end{array}\right)$\tabularnewline
\hline 
\end{tabular}
\par\end{centering}
\caption{Comparison of some physical quantities}
\end{sidewaystable}

\section{Conclusion}

\ ~~\ \ \ ~~~ Hypercomplex division algebra (viz. quaternion)
is regarded as a necessity to study an additional dimension of any
system or theory in a single framework. In this case, quaternonic
representation of four-potential, four-momentum, four-position etc.
have been discussed in view of Minkowski structure. The generalized
quaternionic rotational Dirac equation and its solutions for fermions
without EM-field has been discussed in equation (\ref{eq:29}). The
generalized QRD equation has the benefit of representing both aspects
as rotational energy and rotating momentum corresponding to quaternionic
scalar and vector components. The most important feature of quaternionic
analysis is that it allows for the dual representation of four vectors
in a unified structure. As a result, the EM-field interaction in the
extended QRD equation is stated in equation (\ref{eq:44}), where
equations (\ref{eq:45}) and (\ref{eq:48}) define the resulting quaternionic
rotational EM-energy corresponding to the scalar component and quaternionic
rotational angular momentum corresponding to the vector component,
respectively. Accordingly, the electric and magnetic angular momentum
are established in terms of quaternionic generalized rotational electric
and magnetic vector potentials. Furthermore, the four component Dirac
spinors with spin up and down states have been used to examine the
solutions for quaternionic rotational energy and momentum equations
in the EM-field. The generalized Schrodinger-Pauli-like energy equation,
which is related with unperturbed and perturbed energies as a result
of EM-field interaction, has been constructed in equation (\ref{eq:100}).
Due to EM-interaction in a fermionic field, we have interpreted the
dual perturbed energies, respectively, the generalized electric dipole
energy and the generalized magnetic dipole energy in view of quaternionic
analysis. Table-3 shows a comparison of the linear and rotational
behavior of several quaternionic physical quantities. The beauty of
present analysis, to examine the different symmetries the extended
quaternionic rotational Dirac equation in presence of EM-field has
been shown to be well invariant under the Lorentz, gauge, duality,
and CPT invariances. With the interaction of EM-field, the extended
QRD equation possesses symmetrical structure in electric and magnetic
fields, with two four-potentials associated with Dirac fermions (or
anti-fermions) including the simultaneous existence of electric and
magnetic charge of rotating dyons.

\end{document}